# Combining human and machine learning for morphological analysis of galaxy images

Evan Kuminski,
Lawrence Technological University, 21000 W Ten Mile Rd., Southfield, MI 48075, USA
Email: ekuminski@ltu.edu

Joe George,
Lawrence Technological University, 21000 W Ten Mile Rd., Southfield, MI 48075, USA
Email: jgeorge@ltu.edu

John Wallin,
Middle Tennessee State University, 1301 E Main St, Murfreesboro, TN 37130, USA
Email: john.wallin@mtsu.edu

Lior Shamir,
Lawrence Technological University, 21000 W Ten Mile Rd., Southfield, MI 48075, USA
Email: lshamir@mtu.edu

**Abstract:** The increasing importance of digital sky surveys collecting many millions of galaxy images has reinforced the need for robust methods that can perform morphological analysis of large galaxy image databases. Citizen science initiatives such as Galaxy Zoo showed that large datasets of galaxy images can be analyzed effectively by non-scientist volunteers, but since databases generated by robotic telescopes grow much faster than the processing power of any group of citizen scientists, it is clear that computer analysis is required. Here we propose to use citizen science data for training machine learning systems, and show experimental results demonstrating that machine learning systems can be trained with citizen science data. Our findings show that the performance of machine learning depends on the quality of the data, which can be improved by using samples that have a high degree of agreement between the citizen scientists. The source code of the method is publicly available.

Keywords: Galaxies, Data Analysis and Techniques.

## 1. Introduction

Galaxies have diverse and complex shapes, and their morphology carries fundamental information about the past, present, and future universe. Many morphological schemes of galaxies have been proposed, starting from the broad morphological classification (spiral or elliptical), its class on the Hubble sequence, or more specific morphological features such as the number of spiral arms, number of nuclei, the size of the bulge, etc.

While in the pre-information era galaxies were observed and imaged manually, in the past decade digital sky surveys powered by robotic telescopes have produced very large databases of galaxy images, reinforcing the need for methods that can analyze massive sets of galaxy images. The Sloan Digital Sky Survey (York et al., 2000; Strauss et al., 2002) has imaged several hundred million galaxies so far, and sky surveys such as the Dark Energy Survey (DES) and the Large Synoptic Survey Telescope (LSST) will image billions of galaxies. Since there is no practical way to examine these galaxy images manually, automatic analysis methods will be required to mine for discoveries in these big image data and turn them into knowledge.

Methods for automatic morphological analysis of galaxy images have been proposed, and include GALFIT (Peng, Impey & Rix, 2002), GIM2D (Simard, 1998, 2011; Simard et al., 2011), the Gini coefficient method (Abraham, Van Den Bergh & Nair, 2008), CAS method (Conselice, 2003), and MID statistics (Freeman et al., 2013). However, these methods did not provide a complete solution to the complex problem of automatic morphological analysis of galaxy images, and led to the contention that practical classification of large datasets of galaxy images should be carried out by humans (Lintott et al., 2008, 2011; Willett et al., 2013; Keel et al. 2013). Another approach to analyzing a large number of galaxies is the use of non-expert "citizen scientist" volunteers, who access the galaxy images through a web-based user interface, and submit their annotation of the celestial object to a central database. The successful implementation of that concept led to the Galaxy Zoo project (Lintott et al., 2008), which is part of the Zooniverse citizen science initiative. However, since digital sky surveys such as LSST will acquire billions of galaxy images, citizen science alone will not be able to provide a scalable solution for analyzing databases acquired by future sky surveys.

For instance, in Galaxy Zoo (Willett et al., 2013) ~300K galaxies were analyzed in about three years of work. Even if assuming that all galaxies were analyzed with perfect accuracy by the citizen scientists, LSST is expected to image ~$10^{10}$ galaxies (Borne et al., 2009), which will take the citizen scientists ~$10^6$ years to analyze with the same rate and group size as in Galaxy Zoo 2. Even if the number of well resolved images is 1000 times lower, this task is still beyond the capability of any group of volunteers. Therefore, while citizen science has provided an effective solution for analyzing the morphology of large numbers of galaxies, the increasing data collection power of digital sky surveys reinforces the development of automatic methods that will enhance the manual analysis. These methods can be combined with citizen science analysis of the image data, providing a solution that can scale with extremely large databases of galaxy images (Borne, 2013).

Galaxy images can also be analyzed by using machine learning (Shamir, 2009; Banerji et al., 2010; Huertas-Company et al., 2011). These methods are based on a pattern recognition algorithm trained with manually annotated data, and then the patterns found in these data are used to classify unknown samples that the algorithm was not trained with. By using training data annotated by citizen scientists, effective machine learning algorithms can potentially be trained with large and clean image datasets, ultimately providing a robust solution for analyzing far larger databases of galaxy images.

Here we test the use of galaxy image data classified by citizen scientists as training data for machine learning. The dataset annotated by a large number of human participants can be sufficiently large to train a machine learning system, and the quality of the annotations provides a consistent and clean training set that can be used by machine learning algorithms.

## 2. Galaxy morphology analysis method

Galaxy images are diverse and complex, and therefore require comprehensive image analysis algorithms that measures many different aspects of the visual content. The image analysis method used in this study is Wndchrm (Shamir et al., 2008a), which first extracts a large set of numerical image content descriptors reflecting complex image morphology. The Wndchrm algorithm was developed initially for cell biology (Shamir et al., 2008b), but demonstrated its efficacy for other tasks that require comprehensive morphological analysis such as art (Shamir & Tarakhovsky, 2012; Shamir et al.,2010a). It was also applied to basic tasks in galaxy morphology such as automatic classification between the broad galaxy morphological types of spiral, elliptical, and edge-on (Shamir, 2009), unsupervised analysis of simulated galaxy mergers (Shamir, Holincheck & Wallin, 2013), and automatic detection of peculiar galaxies (Shamir, 2012; Shamir & Wallin, 2014).

In summary, Wndchrm first extracts a large set of 2883 numerical image content descriptors that include texture features such as the Gabor filters and Haralick and Tamura textures, statistical distribution of the pixel intensities such as multi-scale histograms and first four moments, high contrast features such as edge and object statistics, polynomial representation of the pixel values such as Chebyshev statistics and Zernike polynomial, Radon features, fractals, and the Gini coefficient described in (Abraham, Van Den Bergh & Nair, 2003). These features are not extracted merely from the raw pixels, but also from transforms of the raw pixels and transforms of transforms. The image transforms include the Fourier transform, Chebyshev transform, Wavelet transform (Symlet 5), and edge magnitude transform, as well as combinations of these transforms (Shamir et al., 2008a; Shamir et al., 2009; Shamir et al., 2012; Shamir, 2012; Shamir & Tarakhovsky, 2012; Shamir, Holincheck & Wallin, 2013).

Since the Wndchrm scheme is designed for different image classification problems, it can be assumed that not all numerical content descriptors are equally informative for a given image classification problem, and some non-informative numerical content descriptors can also add noise and degrade the classification accuracy. To select the most informative numerical content descriptors each feature is assigned with its Fisher discriminant score, and 95% of the features with the lowest Fisher discriminant scores are rejected. Then, the classification is performed using a Weighted Nearest Neighbor scheme such that the Fisher discriminant scores assigned to the features are used as weights (Shamir et al., 2008a; Shamir et al., 2012; Shamir, 2012; Shamir & Tarakhovsky, 2012; Shamir, Holincheck & Wallin, 2013), as described in Equation 1:

$$1) \quad d(x,c) = \frac{\sum_{t \in T_c} \left[ \sum_{f=1}^{|x|} W_f^2 (x_f - t_f)^2 \right]^p}{|T_c|},$$

where $T_c$ is the training set of class $c$, $t$ is a feature vector from $T_c$, $|x|$ is the length of the feature vector $x$, $x_f$ is the value of image feature $f$, $W_f$ is the Fisher discriminant score of feature $f$, $|T_c|$ is the number of training samples of class $c$, $d(x,c)$ is the computed distance from a given sample $x$ to class $c$, and $p$ is the exponent, which is set to -5 as thoroughly discussed in (Orlov et al., 2008). The comprehensive set of image content descriptors and the selection of the informative image features allow the application of the algorithm to a broad variety of complex image data (Shamir et al., 2008a; Shamir et al., 2012; Shamir, 2012; Shamir & Tarakhovsky, 2012; Shamir, Holincheck & Wallin, 2013; Shamir, 2013). Source code for the method is publicly available (Shamir et al., 2013).

## 3. Galaxy image data

The data used in the experiment are images of SDSS galaxies analyzed by Galaxy Zoo 2 as part of the Zooniverse citizen science initiative (Willett et al., 2013). Each image is a 120x120 JPEG image downloaded from the Catalog Archive Server (CAS) of SDSS. Since Galaxy Zoo 2 galaxies are of different angular sizes, the galaxy images are downloaded such that the first image of each galaxy has a scale of 0.1 arcseconds per pixel. Then, an Otsu binary transform (Otsu, 1971) is applied to separate foreground pixels from the background. If more than 40 foreground pixels are detected on the edge of the image, the scale is increased by 0.05 arcseconds per pixel and the image is downloaded again, until no more than 40 pixels are detected on the edge. That process leads to images that contain the entire galaxy. These images are smaller than the 424x424 images used in Galaxy Zoo 2, but they contain many fewer background pixels than the Galaxy Zoo 2 images. The initial scale of 0.1 arcseconds per pixel was determined empirically as a scale that is too small to contain the entire galaxy in the frame. Galaxy Zoo 2 galaxies have relatively large angular sizes, and therefore the initial size might need to be adjusted when processing images of a set of smaller galaxies.

The Galaxy Zoo 2 data release includes detailed morphological information about 304,122 galaxies annotated by Zooniverse citizen scientists - non-expert volunteers who do not have formal training as scientists, but are able to contribute to scientific research in tasks such as basic data analysis. We used in this study the 245,609 original Galaxy Zoo 2 images, and excluded galaxies from Stripe 82 and other galaxies that were added at a later time. Each citizen scientist was presented with a galaxy image, and a set of questions they needed to answer about its morphology such as signs of spirality, the number of arms of the galaxy, and more, and provided answers to these questions before continuing to the next galaxy image. Each question in the sequence is based on the answer to the previous question (Willett et al., 2013). For instance, if the participant's reply to the first question is that the galaxy is smooth, they were not asked about the number of arms, but instead were asked about the degree of roundness of the object.

This study is focused on galaxy morphology reflected by a set of questions that the Zooniverse participants answered for each galaxy they annotated (Willett et al., 2013). Galaxy Zoo 2 has a total of 11 questions, and each participant provides answers to between one to eight questions determined by the path on the decision tree, such that the answer to a certain question determines the next question in the sequence (Willett et al., 2013).

The morphological features analyzed in the study are summarized in Table 1.

| Question number | Question | Possible answers |
|---|---|---|
| 1 | Is the galaxy simply smooth and rounded, with no sign of a disk? | Smooth, feature or disk, star or artifact |
| 2 | Could this be a disk viewed edge-on? | Yes, no |
| 3 | Is there a sign of a bar feature through the centre of the galaxy? | Yes, no |
| 4 | Is there any sign of a spiral arm pattern? | Yes, no |
| 5 | How prominent is the central bulge, compared just noticeable with the rest of the galaxy? | No bulge, just noticeable, obvious, dominant |
| 6 | Is there anything odd? | Yes, no |
| 7 | How rounded is it? | Completely round, in between, cigar-shaped |
| 8 | Is the odd feature a ring, or is the galaxy disturbed or irregular? | ring, lens or arc, disturbed, irregular, other, merger, dust lane |
| 9 | Does the galaxy have a bulge at its centre? If so, what shape? | rounded, boxy, no bulge |
| 10 | How tightly wound do the spiral arms appear? | Tight, medium, loose |
| 11 | How many spiral arms are there? | 1, 2, 3, 4, more than four, can't tell |

Table 1. The 11 questions of Galaxy Zoo 2 and the possible answers to each question (Willett et al., 2013).

The citizen science answers for each question provided the data for the pattern recognition experiments. For example, the first question provided two classes of images - galaxies that were classified by the Zooniverse citizen scientists as smooth and round, and galaxies that were classified as not smooth and round. In the case of the first question the goal of the machine learning method was to automatically identify between these two classes. That was repeated for each of the questions described above.

As discussed above, GZ2 citizen scientists answer the questions in a sequence such that the next question is determined by their answer to the previous questions. That might lead to some answers that are based on a minority probability and can add confusion to the algorithm. For instance, if 80% of the citizen scientists voting on question 2 of a certain galaxy determine that the galaxy is not an edge-on galaxy, it can be assumed that the galaxy is not edge-on. However, the remaining 20% who annotated the galaxy as edge-on are then prompted to answer about the shape of the bulge (question 9), which is based on a previous answer with minority probability and might therefore not be optimal to be used as ground truth for training a machine learning system.

To reduce the effect of answers based on minority votes, we set a threshold of agreement such that votes that were different from the majority of the votes were ignored in the following questions. For instance, if the agreement threshold is 80% (Bamford et al., 2009; Willett et al., 2013), and more than 80% voted "no" on question 2, the answers to the following questions of those that voted "yes" are ignored.

Each galaxy was classified by multiple citizen scientists (Willett et al., 2013). The median number of classifications for each galaxy was 44, and the minimum number was 16. Supervised machine learning requires clean ground truth data, and therefore the training of a machine learning algorithm is generally more effective when using the cleanest available data. The multiple classifications for each galaxy allow selecting the most consistent subsets of the data by using the degree of agreement between the citizen scientists, such that a higher agreement between the human classifiers reflects a more accurate classification. For instance, it is possible to select a subset of the data such that 90% or more of the voters agree on. If the agreement threshold is set to 90% then only galaxies that 90% or more of the citizen scientists marked as round are selected for the *round* class, and only galaxies that 90% or more of the citizen scientists classified as not round are selected for the *not round* class. The rest of the galaxies are ignored. As mentioned above, the 90% agreement is also required in the sequence of answers that led to the question. For the degree of agreement, two different methods of Galaxy Zoo 2 were used. One is the raw count of the votes for each question, and the other is the correction of the votes for magnitude bias (Willett et al., 2013).

High level of disagreement between the citizen scientists on a certain morphological feature of a certain galaxy indicates that the morphology of that galaxy is not clear, and therefore cannot be used to effectively train a machine learning algorithm. However, the number of samples that satisfy an agreement level threshold decreases as the agreement threshold gets higher, leading to a smaller dataset, consequently leaving less samples that a machine learning system can be trained with. Therefore, optimizing the performance of the machine learning system requires finding the degree of agreement that provides the best classification accuracy for each of the questions.

To balance the dataset used in each experiment we assigned each class with the number of samples equal to the number of samples in the smallest class. For instance, if a certain morphological feature had two classes, and the number of galaxies that satisfied the 90% agreement level was 1000 for the first class and 1500 for the second class, we randomly removed 500 images from the second class in each run so that the two classes would have an equal number of samples. Balancing the number of samples in the classes is important for avoiding bias in the classification (Shamir et al., 2010b), as some machine learning methods can prefer assigning a test sample with classes that are represented by more samples in the training set.

## 4. Machine learning using citizen science data

Figure 1 shows the classification accuracy of the different questions for different levels of agreement among the citizen scientists. For each question the samples are divided into training and test sets such that the number of training and test samples are equal across all classes as described in Section 3, and the classification accuracy is determined by the number of test samples for which the automatic classification was in agreement with the citizen science classification, divided by the total number of test samples.

Each experiment was repeated 40 times such that in each run 90% of the samples were randomly allocated for training, and the remaining samples were used for testing. The classification accuracy of the 40 runs was averaged to provide the final classification accuracy for the question.

As the Figure 1 shows, when considering the human classification as ground truth the automatic method was at least 85% accurate for eight out of 10 morphological features analyzed in Galaxy Zoo 2 when compared to the results of the volunteers. For questions 1, 2, 4, and 7 the classification accuracy was higher than 95%. For some of the questions the number of galaxies was very low or zero when the agreement threshold was high, and therefore some questions do not have results of automatic classification accuracy at some agreement thresholds.

It is also clear from the graph that the classification accuracy increased as the agreement threshold between the citizen scientists gets higher, showing that higher agreement between the citizen scientists provides more consistent data, leading to better ability of the machine learning method to classify between the samples despite the smaller training set. However, when the training set became too small the machine learning became less accurate, resulting in degraded classification accuracy. Table 2 shows the number of galaxies per class and the total number of galaxies used for each of the questions, and for different levels of agreement. As mentioned above, the number of galaxies per class used by the machine learning algorithm was the number of galaxies in the smallest class.

| Question | >50% | >60% | >70% | >80% | >90% | >95% | >97% |
|---|---|---|---|---|---|---|---|
| 1 | 25000 (241679) | 25000 (213890) | 25000 (181715) | 25000 (132748) | 19693 (50946) | 6635 (19248) | 2332 (12193) |
| 2 | 10367 (61515) | 6705 (53058) | 3955 (41193) | 2003 (27891) | 645 (18325) | 225 (10154) | 127 (6761) |
| 3 | 13993 (55488) | 9964 (46176) | 6506 (37181) | 3720 (27788) | 1399 (17646) | 386 (9913) | 171 (6629) |
| 4 | 9846 (55396) | 4334 (42130) | 1522 (32119) | 323 (23681) | 11 (15106) | 1 (8421) | 0 (5479) |
| 5 | 13028 (42780) | 5866 (24385) | 510 (10583) | 110 (2499) | 3 (143) | 0 (9) | 0 (2) |
| 6 | 22889 (242291) | 15791 (224645) | 9921 (202509) | 5369 (170574) | 1957 (115692) | 691 (65638) | 417 (46159) |
| 7 | 24203 (172761) | 16442 (138128) | 8593 (100328) | 2117 (55774) | 103 (9545) | 6 (961) | 4 (221) |
| 8 | 37 (15219) | 9 (7483) | 1 (3078) | 0 (1008) | 0 (163) | 0 (18) | 0 (11) |
| 9 | 97 (9272) | 44 (4860) | 18 (2076) | 8 (562) | 3 (48) | 0 (4) | 0 (1) |
| 10 | 5471 (33536) | 3371 (16289) | 1337 (6332) | 119 (2061) | 6 (469) | 0 (118) | 0 (51) |
| 11 | 226 (21814) | 120 (14966) | 58 (10900) | 24 (8490) | 0 (5948) | 0 (3568) | 0 (2343) |

Table 2. The number of galaxies per class used in Wndchrm (above) and the total number of galaxies (below, in parentheses) for each GZ2 agreement threshold.

In question 1 there was a small number of images that were identified as stars or artifacts. Since for automatic classification the number of samples should be the same for all classes (Shamir et al., 2010b), including the stars or artifacts would force the other classes to also contain just very few samples, and therefore the objects classified as stars or artifacts were ignored. When the number of galaxies was very high, it was limited to 25,000 galaxies per class due to response time considerations. In several cases the number of galaxies per class dropped to zero when the agreement threshold was 80% or 90%, as one of the classes did not have even one sample that satisfied that level of agreement.

For the irregular galaxies (question 8), just a few galaxies met the threshold of 60% of agreement between the human voters of Galaxy Zoo 2, and the classes "lens" and "disturbed" were ignored as they contained no galaxies. Previous studies show that the problem of automatically detecting irregular galaxies can be approached as a novelty detection problem rather than a classification problem (Shamir, 2012; Shamir & Wallin, 2014). It should be noted that the use of citizen science analysis for the detection of irregular galaxies should also be examined carefully, as identifying peculiar celestial objects effectively requires substantial experience that might be beyond the knowledge of the typical non-astronomer.

The method also failed to provide good classification accuracy for question 11, which is the number of arms of the galaxy. When using galaxies that were classified by citizen scientists with agreement threshold of 60%, the computer method identified the number of galaxies with accuracy of just ~34%. These performance figures suggest that the method might not be informative for that specific morphological feature. Automatically determining the number of arms in a galaxy is known as a difficult task even when the data are clean (Davis & Hayes, 2014; Seigar et al., 2005, Peng et al., 2010).

The difficulty in identifying the number of arms can also be explained by the inconsistency of the data, as the degree of disagreement between the answers provided by the citizen scientists to that question was high. For instance, when the agreement threshold between the citizen scientists was set to 60% each class had 120 samples, and when it was set of 75% the number of galaxies per class was 31. However, the experiments with question 9 show that the algorithm is able to achieve higher classification accuracy with a smaller training set, indicating that answering question 11 is more difficult for the machine learning algorithm than answering question 9.

As discussed in Section 3, the experiments were also done using the correction of the raw votes for the magnitude bias. The results of the experiment using the debiased data are displayed by Figure 2.

As the figure shows, using the debiased GZ2 data led to a slight change in the ability of the algorithm to automatically classify the galaxies by their morphology. The questions that improve the most by using the debiased data are question 2 and 4.

The automatic classification of the galaxies is based on a very large feature set of 2883 image features, and the high dimensionality of the feature set makes it difficult to conceptualize the identification of galaxy images based on each question. The most informative features are selected automatically by their ability to provide separation between the classes in the training set, and therefore for each question different image features are selected. The dynamic selection of image features allows the same algorithm to answer the different questions. Table 3 shows the most informative groups of image numerical content descriptors and the image transforms they are extracted from, as determine by the Fisher discriminant scores.

| 1 | 2 | 3 | 4 | 5 | 6 | 7 | 9 | 10 |
|---|---|---|---|---|---|---|---|---|
| Fractal (edge) | Zernike (FFT) | Zernike (edge) | Haralick (Chebyshev+FFT) | Chebyshev (FFT) | Haralick (edge+FFT) | Zernike (raw) | Zernike (raw) | Fractal (edge) |
| Zernike (FFT) | First 4 moments (FFT) | Zernike (FFT) | Fractals (FFT+wavelet) | Zernike (Chebyshev) | First 4 moments (Chebyshev+wavelet) | Zernike (FFT) | Fractal (FFT) | Zernike (FFT) |
| Zernike (raw) | Chebyshev (edge+FFT) | Fractals (edge) | Zernike (raw) | Fractal (FFT) | Haralick (Chebyshev + FFT) | First 4 moments (FFT) | Chebyshev (FFT) | Zernike (FFT+Chebyshev) |
| Haralick (Chebyshev) | First 4 moments (Chebyshev + FFT) | Zernike (edge+wavelet) | Haralick (FFT) | Zernike (raw) | Zernike (Chebyshev) | Zernike (edge+wavelet) | Multiscale histograms (Chebyshev+FFT) | Haralick (Chebyshev) |
| Haralick (edge) | Multiscale histograms (Chebyshev + wavelet) | Haralick (wavelet) | Fractals (raw) | First 4 moments (wavelet+FFT) | Zernike (Chebyshev+FFT) | First 4 moments (FFT) | Haralick (edge+wavelet) | Fractal (FFT) |
| Chebyshev (FFT+wavelet) | Multiscale Histograms (raw) | Haralick (raw) | Haralick (raw) | Haralick (edge+wavelet) | Haralick (edge+FFT) | First 4 moments (raw) | Zernike (Chebyshev) | Haralick (FFT) |

Table 3. The most informative group of features used for each question.

Many of the numerical content descriptors in the table are extracted from image transforms or multi-order image transforms, and are therefore difficult to conceptualize by the human intuition. For

12instance, the Haralick textures extracted from the Fourier transform of the edges of the original image contain information that is useful for classifying the image, but due to the transforms it is difficult to intuitively identify the visual cues that it extracts.

As the table shows, many of the features are informative for more than one question, but the features are different for each question, and no single set of features can be used for all possible questions. The reason for the different image content descriptors is that each morphological feature is better identified by different image cues, and therefore different types of numerical content descriptors are more informative for different galaxy morphological features. For instance, fractal features are the most informative descriptors for the first question, as the spiral shape of a galaxy has a certain degree of fractality, which does not exist in smooth galaxies (Shamir, 2009). For the same reason it can be assumed that fractals are also dominant in the classification based on question 4, which is the existence of signs of spiral arms, and question 3, where fractals can differentiate between bars and regular spiral arms. In question 2, pixel statistics such as multi-scale histogram and first four moments are found informative, which can be explained by the different surface size of a face-on galaxy compared to edge-on. Multi-scale histograms are also useful for question 5, where a prominent bulge is expected to lead to a small group of pixels brighter than the rest of the foreground pixels, and therefore be reflected by the pixel intensity histogram.

A group of numerical content descriptors informative for all questions is the Zernike polynomials. Zernike features are effective for reflecting variations in the unit disk (Teague, 1979), and are therefore useful for the analysis of morphology of round objects such as cells (Shamir et al., 2008b), round joints (Shamir et al., 2009) and in the case of this study also galaxies.

Question 7, which is the roundness of the object, is classified by a combination of Zernike polynomials and pixel statistics, as the Zernike features are sensitive to the unit disk and the pixel statistics is affected by the number of foreground pixels, which grows as the object is more round.

## 5. Conclusion

As robotic telescopes acquiring big astronomical image data become increasingly important, methods that can automatically analyze astronomical images will be required to turn these data into knowledge and optimize the scientific return. Digital sky surveys are becoming increasingly important in astronomy, and that trend is bound to continue. Future surveys such as LSST will produce the world's largest public database, reinforcing the development of automatic methods that can take a galaxy image as input, and provides its basic morphological characteristic as output. Such automatic methods can analyze all galaxy images acquired by LSST, and create data products of galaxy morphological features that will be added to the galaxy photometric data in future LSST data releases.

Automatic methods can also work in concert with manual analysis and citizen science. For instance, citizen science can be effective for tasks that are more difficult for pattern recognition methods, such as detection of peculiar galaxies. Automatic algorithms can mine the database of ~10B galaxy images and detect peculiar galaxy candidates (Shamir, 2012; Shamir & Wallin, 2014), and human analysis can complete the analysis by removing the false positives, which are the vast majority of the data, to detect the actual rare galaxy types. Manual analysis can also be useful to validate the data and produce clean datasets by annotating in-between cases and galaxies that machines cannot classify with high accuracy. In such data analysis pipeline, human analysis will be used after automatic pre-processing of the data. Human analysis can be more accurate than machine analysis, but since the availability of human analysis is limited by the number of citizen scientists, its effectiveness can be optimized if used to analyze data that cannot be analyzed effectively by computers.

Another important link between human and machine analysis is the development of pattern recognition systems that can analyze galaxy images. Manually annotated datasets with sample size of a few million samples can be used to select clean subsets of samples, and train machine learning systems that can then automatically perform the data annotations. The superior pattern recognition of the human brain utilized by citizen science can also be tasked with doing routine checks of automatically classified data and examining cases that are unusual or ambiguous. A partnership between human computing and automated algorithms may ultimately be one of the best approaches for dealing with image classification in large astronomical data sets.

One of the interesting aspects of citizen science data is the level of agreement between the voters for each image. With a single expert, only one classification for each image is provided. However, when a few dozen volunteers examine the images the level of agreement depends on the characteristics of the image being examined. Supervised machine learning algorithms require consistent ground truth data for training, and the results of this paper show strong link between the agreement of the citizen scientists on the data annotation and the ability of the machine learning system to analyze these data. By leveraging the range of votes we obtain a measure of reliability of the classifications that cannot be achieved by a single expert, providing a clean set of samples. This measure of reliability in the human classifications provides a natural match for machine learning algorithms, which often provide a likelihood value for each classification.

Another interesting aspect of citizen science data is how some classification questions seem to be harder than others. Asking volunteers to determine if a galaxy is round or has features seems to be relatively simple task that results in high agreement between volunteers at least in some images. However asking volunteers to count the number of spiral arms results in lower consistency data. For this particular question, this inconsistency may be because few galaxies have visible spiral arms and even fewer have an unusual number of spiral arms. These questions may help identify some cases with unusual characteristics such as 3-armed spirals, but it would be difficult to draw any strong statistical conclusions from the crowd-sourced results. Knowing that these questions are particularly difficult for volunteers suggests that some automated approaches that might provide better results in these particular cases should be developed and applied.

Automatic analysis of morphological features of galaxies can be conceptualized as a pure classification problem, such as the broad classification between elliptical and spiral galaxies. However, galaxy morphologies result from a complex set of interactions through hierarchical assembly, there are many in-between cases, and therefore the classification of galaxies to one of several distinct classes can be oversimplification of the problem. Similar to citizen science, a machine learning approach to the problem can be assigning each galaxy a set of likelihood values to each morphological class, in addition to the most likely distinct class in which the celestial object belongs (Huertas-Company et al., 2011). These likelihood values can reflect the uncertainty of the classification.

For the training of a machine learning system, it is important to use consistent data. Since datasets of celestial objects generated by digital sky surveys are normally large, in many studies the in-between cases can be ignored when training a machine learning system, while the remaining dataset of "clean" samples can still be sufficiently large for effective training.

One of the obvious disadvantages of the method is the time and efforts involved in collecting human annotation of hundreds of thousands of celestial objects, each is annotated by multiple voters to obtain a set of morphological descriptors. However, the successful experience of Zooniverse has clearly shown that analyzing such datasets manually is feasible, and citizen science data makes it possible to quantify the accuracy of machine learning algorithms.

17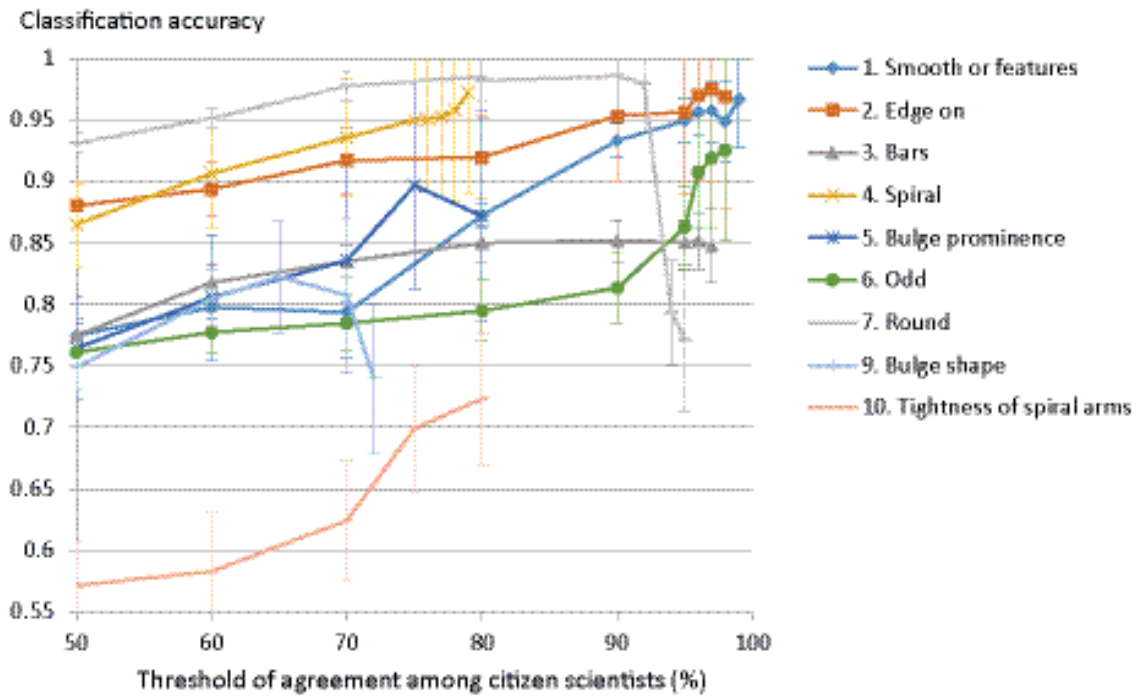

Figure 1. The classification accuracy of the automated method for the different questions and different agreement thresholds among the citizen scientists who annotated the data. The classification accuracy is measured by the number of test samples that their automatic classification was in agreement with the citizen science classification, divided by the total number of test samples. The error bars show the standard deviation of the classification accuracy in the 40 different runs such that in each run different samples are randomly allocated to training and test sets.

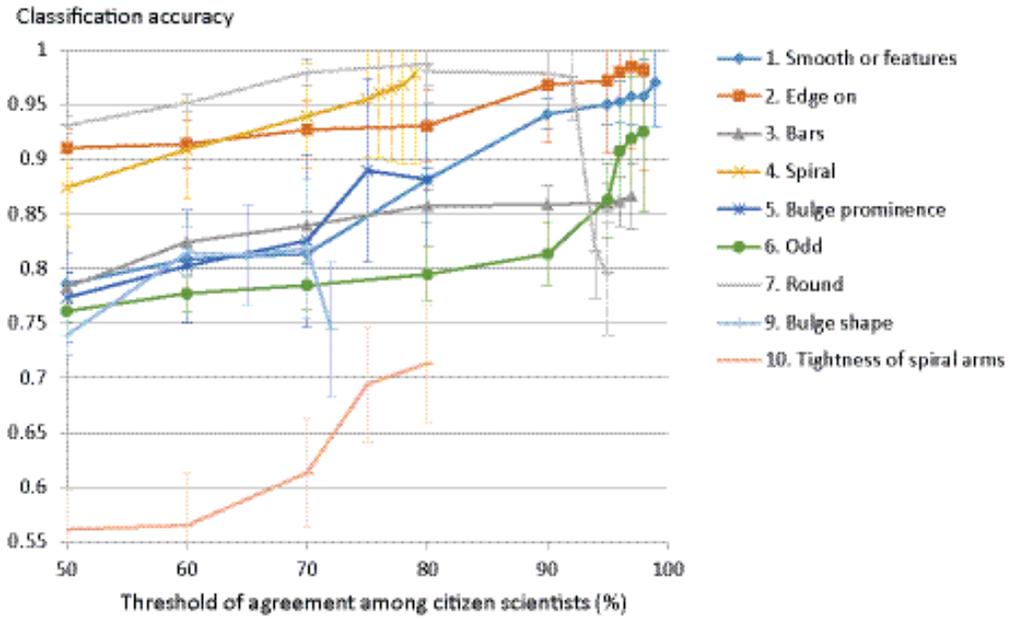

Figure 2. The classification accuracy of the automated method for the different questions and different agreement thresholds using Galaxy Zoo 2 debiased data.